\documentclass[aps,onecolumn,preprint,prd,superscriptaddress,nofootinbib]{revtex4}
\usepackage{graphicx}
\usepackage{amsmath,amssymb,color,mathrsfs,verbatim,psfig,bbm,wasysym, pstricks,epsfig}

\newcommand{\beq}{\begin{equation}}
\newcommand{\eeq}{\end{equation}}
\newcommand{\beqa}{\begin{eqnarray}}
\newcommand{\eeqa}{\end{eqnarray}}
\newcommand{\ba}{\begin{array}}
\newcommand{\ea}{\end{array}}
\newcommand{\mo}{\mathcal{O}}
\newcommand{\ml}{\mathcal{L}}

\begin{document}

\begin{titlepage}
 
\thispagestyle{empty}

\vspace{7.5cm}
\begin{center} 
\vskip 7.5cm
\phantom{A}
\vspace{2.7cm}
{\LARGE \bf  Implications of a New Light Scalar\\ \vspace{0.27cm}
 Near the Bottomonium Regime}
\vskip .7cm

\vskip .7cm
{\large Matthew Baumgart$^{1}$ and Andrey Katz$^{2}$}\vskip 0.4cm
{\it $^1$Department of Physics and Astronomy, Johns Hopkins University, Baltimore, MD 21218}\\
{\it $^2$Center for the Fundamental Laws of Nature, Jefferson Physical Laboratory,\\ Harvard University, Cambridge, MA 02138}\\
\vskip 1.7cm
\end{center}

\noindent 
We study the decay modes of a new, light spin-0 particle, arguing that if the mass of the (pseudo)scalar is $\sim 11-15$~GeV, it can have an appreciable 
branching ratio into bottomonium, in particular the rare $\eta_b$s.  
Using non-relativistic QCD (NRQCD), we calculate its decay rate to bottomonia for mass splittings greater than the typical momentum transfer within the bound state. 
It can exceed that of decays to other Standard Model fermions
under the assumption of couplings proportional to those of the Standard Model.  
At smaller splittings, where our computational methods break down, we estimate the rate into bottomonia using data-driven methods.
When the spin-0 state decays to bottomonia whose mass is too light to produce $B$-meson 
pairs, we get a qualitatively new experimental signature, decays to $b$-quarks invisible to $b$-tagging.  Such a light, spinless particle can arise in extended Higgs
sectors, making this channel potentially observable in decay chains initiated by the subdominant decay of a Standard Model-like Higgs to a pair of them.

\end{titlepage}

\section{Introduction}
\label{sec:intro}

The LHC will soon cover the entire light Standard Model (SM) Higgs region. There is already convincing evidence that both Atlas and CMS are standing on the edge of 
discovery of a Higgs-like particle with mass close to 125 GeV~\cite{ATLAS:2012ae,Chatrchyan:2012tw}. However, even this excess is confirmed as a new particle, 
many fundamental questions about electroweak symmetry breaking will remain unanswered.  It will then be important to perform precision measurements, which would 
confirm or reject the hypothesis of the SM Higgs, probing various decay channels 
and the total width. Another crucial question to address will be whether it is a standalone particle, or a part of much bigger Higgs sector, as predicted by various beyond 
the SM (BSM) models, {\it e.g.} SUSY, or Higgs as a pseudo-Goldstone boson (pGB). 
 
Certain extended Higgs sectors allow for particles significantly below the weak-scale.  
Usually, it is a pGB of a new, approximate 
global symmetry. One can find an illustrative example in the Next-to-Minimal Supersymmetric Standard Model (NMSSM), which has an approximate R-symmetry in the limit of vanishing gaugino masses and $A$-terms. Therefore, if the
NMSSM has small $A$-terms, it naturally contains a pGB of the continuous R-symmetry~\cite{Dobrescu:2000yn}.  Additionally, little Higgs theories provide an example where the entire Higgs sector arises as the pseudo-Goldstones of 
a spontaneously broken global symmetry (see \cite{ArkaniHamed:2001nc} for an early incarnation and \cite{Schmaltz:2005ky} for a review and further references).

We take this new, light particle, $a$, as part of the Higgs sector, and therefore expect its couplings to be roughly proportional to the SM Yukawas. Therefore, if $m_a > 2m_b$, its dominant decay mode would likely be $a \to b\bar b$, while for $m_a < 2m_b$, it will go 
mostly via $a \to \tau^+ \tau^-$ (however, more exotic scenarios have been considered, {\rm e.g.}~$a \to c\bar c$ \cite{Bellazzini:2009kw} and $a\to gg$ \cite{Bellazzini:2009xt}).
In this paper, we emphasize another non-standard decay mode, which was largely neglected in previous studies (see however a discussion in~\cite{Englert:2011us}). 
We consider $m_a$ large enough to evade the constraints from decays of bottomonia at CLEO and BaBar ($m_a \gtrsim$ 9 GeV \cite{Dermisek:2010mg}), but not too heavy, such that direct decays of the (pseudo)scalar into 
bottomonia might still have appreciable branching fraction. As we will see, this corresponds to the mass region around $\sim 11 - 15$ GeV.
We present a scenario where an important decay mode of a (pseudo)scalar is $a \rightarrow {\rm bottomonium} + X$, 
where the $S$-wave pseudoscalars $\eta_b(n)$, $^3P_J$ states $\chi_{bJ}(n)$, or $^1D_2(n)$ are the principle bottomonia involved ($n$ denotes the various radial 
eigenstates) and $X$ are light hadrons.  
We also show, that if $a$ is a \emph{scalar, rather than a pseudoscalar}, this rate is suppressed. This suppression is less than one order of magnitude and therefore may be difficult to observe.  Nonetheless, a discovery of the mode $a \to {\rm bottomonium} + X
$ might serve as a hint of the parity of $a$.\footnote{For current constraints on a light spin zero-particle see~\cite{Dermisek:2009fd,Dermisek:2010mg,ATLASlighta}. We see that this possibility is wide open in the relevant mass range if the Standard Model-like Higgs is not too light.}  

The collider physics of, say, $a \to \eta_b + X$ decays is challenging, but interesting. 
First of all, the $a$ can appear through non-standard decays of the SM-like Higgs boson. Theoretically it is even possible to achieve a spectrum, where $h \to aa$ is a 
dominant Higgs decay mode, getting the so-called ``hidden Higgs'' (see~\cite{Chang:2008cw,Dermisek:2010tg} for review and references therein). Given very strong hints for a SM-like Higgs at 125 GeV, this scenario is somewhat disfavored.  Nonetheless, 
it is still possible that the process $h\to aa$ is present, but subdominant to the standard $h \to b\bar b $ and $h \to WW^*$ decays.  In fact, if the observed excess at 125 GeV turns out be a SM-like Higgs, then the gluon fusion 
cross section in the SM is $\sim$20 pb, and so with 20 fb$^{-1}$ of data, one would have 400,000 Higgses.  An $\mo(1-10\%)$ branching ratio of $h \rightarrow aa$ would therefore contain thousands of BSM events.  
If one were to observe additional heavy particles from the Higgs sector, other cascade channels would also open.
For example, in the NMSSM, one can produce the $a$ in abundance through heavy charged Higgs decays~\cite{Drees:1999sb} or in association with charginos and neutralinos~\cite{Arhrib:2006sx,Cheung:2008rh}.
Alternatively, one can also look for $a$s radiated from heavy flavors in $b\bar b $ and $t \bar t$ events~\cite{Graham:2006tr}.\footnote{An alternative proposal to look 
for signs of the $a$ is to study the $\eta_b$ masses in detail to look for deviations due to mixing between the two.
Additionally, one can look at the $\tau$-branching ratios of the $\Upsilon$, attributing discrepancies with flavor universality to radiative  $\Upsilon \to \eta_b \gamma$ transitions 
and subsequent $\eta_b$ decays~\cite{Domingo:2008rr,Domingo:2009tb,Rashed:2010jp}.  We will have much to say below about the interaction between the $a$ and 
the $\eta_b$.  However, given the observational rareness of the 
$\eta_b$ and the theoretical uncertainties on their properties, it may well be that the $a$ will allow us to measure properties of the $\eta_b$ rather than the other way 
around.}

Quarkonia decay promptly and usually without leptons in the final state. Therefore, for those bottomonia that are below the two-$B$-meson threshold, the $a$ ends up in 
two jets 
which cannot be $b$-tagged. If the $a$ comes from a rare Higgs decay, these jets can be collimated, resembling the situation in the ``buried Higgs'' \cite
{Bellazzini:2009xt}.  Although the bottomonium decay mode is challenging, it has handles that make it different from a regular QCD jet, which we will elaborate 
in Section \ref{sec:disc}.  The detailed analysis of rare Higgs decays in this channel, though, is beyond the scope of this project.

The layout of the paper is as follows. 
In the next section, we give an introduction to NRQCD and how we can use it to calculate bottomonium production rates. We also discuss a range of validity of this 
technique and show that the decay $a \to \eta_b + X$ is non-negligible in this regime. 
In Section \ref{sec:binding}, we discuss small mass splittings between the $a$ and the bottomonium, where our NRQCD calculation is invalid.  
We estimate the order of magnitude of the $a$'s decay rate to bound states through a particular, computable channel involving $a-\eta_b$ mixing.  We also incorporate
mixing into a computation of the rate for $a \to gg$ over our range of interest. 
Readers wishing to skip to the punchline can look at Figs.~(\ref{fig:totalrates}) and (\ref{fig:br}) for a final plot of decay modes and their branching
ratios, showing where $a \rightarrow {\rm bottomonium} + X$ can be important.  Finally, in the fourth section we discuss our results and conclude. A discussion of open-flavor decays to $b$s in the mixing regime is relegated to the Appendix.

\section{NRQCD calculation}
\label{sec:nrqcd}

For mass splittings between the light spin-0 particle, $a$, and the bottomonium greater than the typical bound quark momentum, ($p_b \sim m_b v$, where $v$ is the quark velocity in the hadronic rest frame)
we can compute the decay $a \rightarrow {\rm bottomonium} \, +X$ in NRQCD simply.  Due to the factorization of 
long from short-distance physics, the calculation combines a perturbative QCD process with a nonperturbative factor taken from data or estimated through its power counting.  After giving an overview of the effective field theory, we present
each part of the computation in turn.  

\subsection{NRQCD basics and validity range}
\label{NRQCDreview}

Non-relativistic QCD (NRQCD) is a rich example of an effective field theory where certain particles ($c$ and $b$-quarks) are light enough to exist as propagating degrees of freedom, but heavy enough that we can integrate out their pair creation,
decoupling quark from antiquark, and can perturbatively expand in powers of their hadronic rest-frame velocity, $v$, as well as their inverse mass, $1/M$,  (we follow the treatment in Ref.~\cite{Bodwin:1994jh}).  Thus, we can work in the non-relativistic limit for the heavy quark fields, giving the following lagrangian,
\beq
\ml_{\rm NRQCD} = \ml_{\rm heavy} + \ml_{\rm light} + \delta \ml,\
\label{eq:nrqcdlag}
\eeq
where $\ml_{\rm light}$ is the usual lagrangian for gluons and light quarks, $\delta \ml$ contains the correction terms that systematically give back full QCD, and 
\beq
\ml_{\rm heavy} = \psi^\dag \left( i D_t + \frac{\bf  D^2}{2M}  \right) \psi + \chi^\dag \left( i D_t - \frac{\bf  D^2}{2M}  \right) \chi.
\label{eq:heavylag}
\eeq
We have separated particle from antiparticle, and our fermions have been reduced to two-component Pauli spinors, with $\psi$ 
annihilating heavy quarks and $\chi$ creating heavy antiquarks.  The main advantages to using NRQCD for computing decays to bottomonia are two-fold: factorization and power counting.  
In fact, as we will show, we will need very little of the full NRQCD machinery once we understand how to make use of these two properties.

The entire effective field theory approach is applicable because we have three parametrically separated scales in the problem:
\begin{itemize}
 \item The heavy quark scale, $M$.
 \item The scale of momentum transfer inside the quarkonium, $M v$. The size of the quarkonium is also characterized by this scale, $r \sim (Mv)^{-1}$.
 \item The quark kinetic energy scale $Mv^2$, which is also an energy splitting between the radial excitations of quarkonium. 
\end{itemize}
For $b$-quarks the typical velocity of the quark in the bound state is $v \approx 0.3$~\cite{Quigg:1979vr}.

The parametric separation between the quark mass and the bottomonium energy and momentum scales
lets us factorize its production into a short-distance computation, where we can work with perturbative QCD, and a long-distance, nonperturbative part, which takes
the form of an expectation value of an NRQCD operator and accounts for the binding into a hadron.\footnote{The validity of factorization for the production of quarkonium in NRQCD has never been proven and 
there is some evidence that the nonperturbative matrix elements are not strictly universal \cite{alTalk,Fleming:2012wy}.  Nonetheless, NRQCD has had quantitative success in calculating production rates 
({\it e.g.~}\cite{Bodwin:1992qr}).  Since we will be taking estimates for our matrix elements of interest anyway, these concerns are beyond the order to which we are working.}
For the case of a particle, $a$, decaying into bottomonium with a modest mass splitting between them, the leading contribution is:
\beq
\Gamma[a \rightarrow H + X] = \sum_n \hat{\Gamma}[a \rightarrow b \bar b (n,{\bf 8}) +g] \langle  \mo^H_n \rangle,
\label{eq:factor}
\eeq
where $n$ labels the angular momentum $(^{2S+1}L_J)$ of the $b \bar b$ pair, ${\bf 8}$ denotes the color state of the $b \bar b$ pair and $X$ is whatever hadrons emerge from the gluon emission as well as the conversion of the color octet $b \bar b$ state 
into a singlet bottomonium.  Additional contributions, which are higher order in perturbative $\alpha_s$ and have 3-body suppression, 
but can get kinematic enhancements can be important for large mass splittings (see~\cite{Cheung:1995ka} for such an effect in $Z$ decays). 
The $\hat{\Gamma}$ term is a straightforward partonic calculation, but we need some way to determine $\langle  \mo^H_n \rangle$, which includes all 
nonperturbative effects. In some cases, $\langle \mo^H_n \rangle$ are directly extracted from experiment, but not all matrix elements have been measured.
In fact, none of those relevant for the $a$ decay are known.  Therefore, in Section \ref{sec:calcNRQCD}, we will use simple power-counting and comparison to measured charmonium rates to estimate those of interest.

\begin{figure}[ht]
\centering
\includegraphics[width=0.8\textwidth]{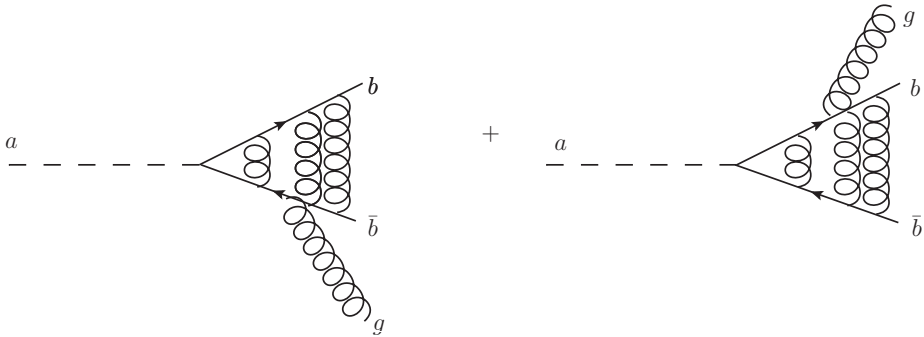} 
\caption{The decay of an $a$ to $b$-quarks with mass splitting $\lesssim \mo(m_b v)$.  In addition to emitting radiation, the $b$s exchange gluons that will bind them into a bottomonium.  In this region of phase space, we cannot completely separate off the binding into the nonperturbative matrix elements of a small number
of operators, but must consider the exchange of multiple gluons coming from infinite towers of operators.}
\label{fig:boundemit}
\end{figure}

Before proceeding to the details of our NRQCD calculation, let us comment on its range of validity.  Our perturbative calculation of $\hat \Gamma$ in Eq.~\ref{eq:factor} assumes that one emits a gluon
harder than those exchanged between the bound quarks.  Otherwise, one no longer has the separation in scales necessary to factorize the effects of binding into an overall multiplicative factor ({\it cf.~}Fig.~\ref{fig:boundemit}).  
Momentum of the softer, ``potential'' gluons is $\mo(m_b v)$.   Therefore, we should demand
\beq
m_{a} - m_{{\rm onium}} \sim p(g) > m_b v~.
\label{eq:limit}
\eeq
Numerically,  $m_b v \approx 1.4 \ {\rm GeV}$ and so we use the NRQCD rate given by Eq.~\ref{eq:factor} 
for splittings greater than around this value. 
Below this scale, the parametric expansion in powers of $v$ breaks down, and one needs to sum infinite towers of operators in order to compute.

\subsection{Calculation}
\label{sec:calcNRQCD}

We take as our starting point an augmented Higgs sector containing a light (pseudo)scalar, 
$a$, with $m_a \gtrsim $ 11 GeV.  This means its splitting with respect to the lightest bottomonium, the $\eta_b(1)$, 
is greater than 1.4 GeV, where we begin to trust the results of perturbative NRQCD ({\it cf.~}Eq.~\ref{eq:limit}).  
We now consider the perturbative and nonperturbative contributions to the $a$ decay in NRQCD in turn.

\subsubsection{Perturbative portion}
\label{subsubsec:pert}

We can have scalar and pseudoscalar couplings of $a$ to $b$-quarks,
\beqa
\ml_{a \bar b b} & = & y_b a \, \bar b b ,\, {\rm or} \\ \nonumber
\ml_{a \bar b b} & = & i \, y_b a \, \bar b\, \gamma^5 \, b~,
\label{eq:pscoupling}
\eeqa
and similarly for other SM fermions.
The origin of $a$ in the Higgs sector motivates its having a large coupling
to the third generation. Strictly speaking, we assume that the $a$'s interactions are 
controlled by the SM Yukawas, and therefore other modes, 
like $c \bar c$, $\tau^+ \tau^-$ are also allowed.  We 
take these channels into account in our rate and branching ratio plots (Figs.~\ref{fig:totalrates} and \ref{fig:br}).\footnote{Our use 
of 
these couplings is not an endorsement of a specific model, but rather an attempt
to determine at the $\mo(1)$ level how branching ratios from the Higgs sector could 
appear.  Specific scenarios ({\it e.g.} large $\tan (\beta)$ NMSSM reducing $c \bar c$) 
can suppress or enhance the rates we depict.  However, generic scenarios minimally affect the relative ratios between them.} 
The large coupling of third generation fermions gives us, along with the 
usual, open-flavor $b \bar b$ decays, $a \rightarrow {\rm bottomonium}\, + X$. This latter decay 
mode is always present, even if it is highly suppressed.
However, it can become an important subleading process if 
the $m_a$ becomes close to the mass threshold of two $B$ mesons. 
\begin{figure}[t]
\centering
\includegraphics[width=0.8\textwidth]{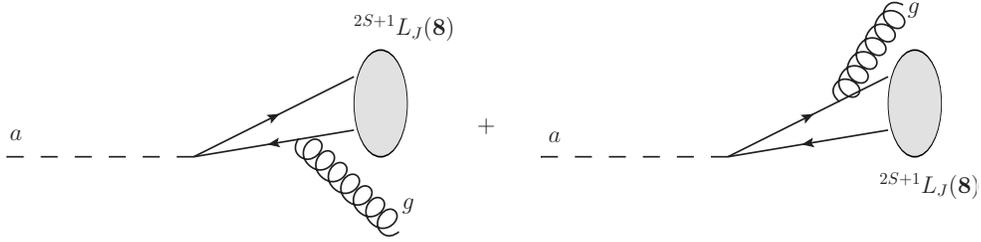} 
\caption{Leading order diagrams for $a \rightarrow {\rm bottomonia} + X$.  Subsequent nonperturbative gluon emissions neutralize the color in both the perturbatively emitted gluon as well as the color octet $b \bar b$ pair.  Unlike the situation depicted in Fig.~\ref{fig:boundemit}, the effects of binding factorize from those of 
radiation, and the former are captured numerically by expectation values of NRQCD operators, which appear as multiplicative factors ({\rm cf.}~Eq.~\ref{eq:factor}).}
\label{fig:emit}
\end{figure}

We first calculate the perturbative decay rates of a (pseudo)scalar
$a$ into the octet $b \bar b $ state and one gluon (see Eq.~\ref{eq:factor} and Fig.~\ref{fig:emit}). 
Despite the velocity suppression that arises from projecting the octet quark combination into the singlet, physical hadron, this 
is the dominant contribution for decays of a light (psuedo)scalar.\footnote{If in a mild abuse of notation we refer to this leading process as $a \rightarrow$ bottomonium $+ g$, then other important processes are 
$a \rightarrow$ bottomonium $+\, b \bar b$, $a \rightarrow$ bottomonium $+\, gg$, and $a \rightarrow$ bottomonium $+\, q \bar q$, where $q$ are light quarks, if we have sufficient phase space for the final state.  For the decay of the $Z$ into charmonium, 
the $\psi q \bar q$ and $\psi c \bar c$ are actually dominant, receiving a kinematic enhancement \cite{Cheung:1995ka}.} 
After projecting the $b \bar b$ octets onto states with well defined angular momentum, we 
find that the $(L=0,\ S= 0)$ and $(L =1,\ S=1)$ modes vanish. 
One can explain this behavior with simple C-invariance.
This may be somewhat surprising as neither the color-octet $b \bar b$ pair nor the gluon is a C-eigenstate. However, the leading order 
diagram (Fig.~\ref{fig:emit}) is identical to a process in QED, with color factorized off as an overall factor, and thus we have an accidental C-symmetry at tree level.  
Therefore, only those processes that would be allowed for $b \bar b$ singlets and {\it photons} will occur.  
We determine the final state C-eigenvalue {\it as if} the constituents were themselves singlets, with each gluon bringing a factor of $-1$, and each 
$b \bar b$ pair a $(-1)^{L+S}$. Since any spin-0 particle is C-even, this leads to a selection rule 
\beq
L+S = {\rm odd}~,
\label{eq:selection}
\eeq
thus ruling out $^1S_0$ and $^3P_J$  $b \bar b$ states as perturbative decay products.

The nonvanishing modes have non-uniform dependence on the mass of the (pseudo)scalar. Defining for 
simplicity 
\beq
\xi \equiv \frac{m_{{\rm onium }}^2}{m_a^2} ~,
\eeq
we get
\beqa
\Gamma({\rm scalar}) & \propto & 1-\xi\\
\Gamma({\rm pseudo}, L=0, S=1) & \propto & 1-\xi\\
\Gamma({\rm pseudo}, L=1, S=0) & \propto & \frac{(1+\xi)^2}{1-\xi}
\eeqa 
One thing to note is that the $^1P_1$ pseudoscalar decay channel grows as 
$m_a$ approaches $m_{{\rm onium}}$ while the other channels vanish. 
The origin of this structure is ultimately in the soft singularity of gauge boson 
emission.  
As we are expanding our $b \bar b$ pair in ``spectroscopic'' states, only those that
do not require relative linear or angular momentum on the part of the gluon can 
get enhanced.  The former constraint rules out having a pole for $S$-wave 
$b \bar b$ states, since we require the quark pair to have identical momenta, 
and thus the gluon must recoil with finite momentum.  The angular momentum 
constraint lets us
see that enhancement for $P$-wave can only occur for a pseudoscalar $a$.  
The $P$-wave state allowed by C-invariance is $^1P_1$.  
For a three-body system, a given orbital angular
momentum eigenstate has parity
\beq
P_{3-{\rm body}} = P_1 P_2 P_3 (-1)^{\ell}(-1)^L,
\eeq 
where in our case $\ell$ is the orbital angular momentum of the $b \bar b$ pair, 
and $L$ that of the gluon with respect to the center of mass of the pair.  The $P_i$ are 
the intrinsic parities of the decay products.
A pseudoscalar allows $L=0$, while a scalar does not.  It may concern the reader that we 
have a pole in our formula associated with an IR divergence.  However, in the regions of 
phase space where we use NRQCD, $\Delta m \gtrsim$ 1.4 GeV, and the $(1-\xi)^{-1}$ factor only
gives a modest enhancement, $\sim 3-4$.  Long before the decay rate can blow up due to this term,
one must account for both the exchange of ``potential'' gluons that bind the fermions as well as 
the reduction of possible hadronic final states from $\Lambda_{QCD}$.  We estimate the decay
rates for mass splittings which must take account of these effects in Section \ref{sec:binding} and find 
that the rate indeed turns over instead of blowing up.

At last we present the explicit results of our calculations. The decay rate that we 
find for the pseudoscalar is
\beq
\label{eq:pseudorate}
\Gamma (a \to {\rm i} + X) = \frac{32\, \alpha_s y_b^2}{m_a m_{{\rm i}}^3}
\left( (1-\xi)\, m_{{\rm i}}^2 \langle \mo_8^{{\rm i}}(^3S_1)\rangle 
+ 4 \frac{(1+\xi)^2}{1-\xi} \langle \mo_8^{{\rm i}}(^1P_1) \rangle \right), 
\eeq 
where i stands for an arbitrary bottomonium state, which we will further specify in the 
next subsection.  We will also discuss systematically why these matrix elements are the leading 
contributions in the $m_b,\, v$-expansions.

The analogous result for a P-even scalar reads
\beq
\Gamma(a_{\rm P-even}\to {\rm i} + X) = \frac{32\, \alpha_s y^2_b}{m_a m_{{\rm i}}^3} 
(1-\xi) \left[ m_{{\rm i}}^2 \langle  \mo^{{\rm i}}_8 (^3S_1) \rangle  + 
4\, \langle  \mo^{{\rm i}}_8 (^1P_1) \rangle \right].
\label{eq:scalarrate}
\eeq   
Even though this result is suppressed compared to the pseudoscalar, it is nonetheless interesting as 
it is part of the bottomonium decay rate of the \emph{Standard Model Higgs 
particle}.  Just as \cite{Cheung:1995ka} found with the $Z$ though, for a 125 GeV Higgs, it will be subdominant by a couple orders of magnitude 
to processes where a $b$ or light quark fragments into a bottomonium, $h \to ({\rm bottomonium}\, +b \bar b)$ or $h \to ({\rm bottomonium}\, +q \bar q)$.
While difficult to observe at the LHC, these processes potentially be measured in future linear colliders,
allowing us to quantify the nonperturbative physics involved.

\subsubsection{Nonperturbative portion}
\label{subsubsec:nonpert}

Even though Eqs.~\eqref{eq:pseudorate} 
and~\eqref{eq:scalarrate} are exact at leading order in QCD and are the lowest-order contributions in the $m_b$ and $v$-expansions of NRQCD,
it will be impossible to translate them to quantative predictions if we know nothing
about the long-distance physics incorporated in the expectation values, $\langle \mo \rangle $.  

For pseudoscalar decay (Eq.~\ref{eq:pseudorate}), the perturbative portion of the $^1P_1$ term is relatively large, so
we ask which physical bottomonium has the biggest 
overlap with the $|b\bar b(^1P_1),{\bf 8} \rangle $. One example is pseudoscalars, namely $\eta_b(n)$. 
As with any other quarkonium state, $|\eta_b \rangle $ is a superposition 
of Fock states, including $b \bar b$ pairs and possibly gluons. Adding a gluon or 
flipping a spin suppresses the probability to find that state by additional powers of 
$v$. Here we expand $|\eta_b \rangle $:
\beqa \nonumber 
|\eta_b \rangle & = & \left| b\bar b (^1S_0)_1\right \rangle + 
\mo (v) \left| b \bar b(^1P_1)_8 \, g\right \rangle +
\mo\left(v^{3/2} \right) \left| b \bar b (^3S_1)_8 \, g  \right \rangle 
 + \, \mo\left( v^2 \right) \left| b \bar b(^1S_0)_{8/1} \, gg \right \rangle \\ 
  \label{eq:vexp}
 && +\, \mo \left( v^2 \right) \left| b \bar b (^1D_2)_{8/1} \, gg \right \rangle + 
\ldots ~,
\eeqa
using the power counting of \cite{Bodwin:1994jh}.\footnote{This expansion closely follows a similar expansion for $Z$ decaying into $\Upsilon$ \cite{Cheung:1995ka}, but uses the slightly different power counting assignment of~\cite{Bodwin:1994jh}.}  
Each electric gluon in the state brings with it 
a factor of $v$, while the magnetic gluon (found in kets where the quarks are spin-flipped) has an additional 
suppression, for an overall factor of $v^{3/2}$.

The $\eta_b$ thus includes the $|b \bar b(^1P_1),{\bf 8}\rangle $ Fock state at $\mo(v)$, which is the lowest possible order the physical singlet can mix with the octet fermion pair, including one dynamical gluon to get an overall singlet. 
Therefore, $a\to \eta_b + X$ is a dominant bottomonium decay mode and we will be interested in the value of $\langle \mo_8^{\eta_b} (^1P_1) \rangle$ to estimate its  rate. Unfortunately, $\eta_b $ is not a well observed particle
and we cannot take the relevant matrix element from data.  However, we can make use of the power counting of NRQCD to 
estimate the size of nonperturbative contributions with factors of $m_b$ and $v$. As a cross-check, we can also take measured values from the charmonium system and rescale them appropriately for bottomonium.  

We begin by understanding how it is that this matrix element describes the process of turning the pair of $b$-quarks into an $\eta_b$.  The operator $\mo^{\eta_b}_8 (^1P_1)$ creates a $b \bar b$ pair, computes its overlap with the 
asymptotic hadronic state $\langle \eta_b + X |$ and takes the outer product with the complex conjugate state so that its vacuum matrix element gives the probability to produce the $\eta_b$ (plus whatever else) from the heavy quark pair.  
We have four fermion fields, which places this operator in $\delta \ml$, in terms of Eq.~\ref{eq:nrqcdlag}.  This makes sense as the creation of $b \bar b$ pairs is short distance physics from the perspective of NRQCD.  From~\cite{Bodwin:1994jh},  we have
\beq
\mo^{\eta_b}_8 (^1P_1) = \chi^\dag \left( -\frac{i}{2} \overleftrightarrow{D}^i \right) t^a \psi \left( a^\dag_{\eta_b} a_{\eta_b}  \right) \psi^\dag \left( -\frac{i}{2} \overleftrightarrow{D}^i \right) t^a\chi,
\label{eq:fourfermi}
\eeq
where
\beq
\left( a^\dag_{\eta_b} a_{\eta_b}  \right) = \sum_X |\eta_b + X \rangle \langle \eta_b + X |.
\label{eq:operator}
\eeq
We are now in a position to estimate the size of the matrix element, taking the state and operator normalization of~\cite{Bodwin:1994jh,Cho:1995vh}.  The size of its
contribution is $m_b^5 v^7$, the breakdown for which is given in Table \ref{tbl:pc}.
\begin{table}[ht]
\begin{tabular}{|l|l|} 
\hline 
Factor & Origin  \\ \hline
$(m_b v)^{-3}$ & Volume factor from\\
 		     & operator spatial integral \\ \hline
 $(m_b v)^6$      & 4 heavy quark fields \\ \hline
 $v^2$		     & Overlap of $b \bar b (^1P_1)_8$ \\
	     & with $\eta_b$ Fock state ({\it cf.} Eq.~\ref{eq:vexp}) \\ \hline
 $(m_b v)^2$      &	$D^i$ in operator \\ \hline\hline
 $m_b^5 v^7$     & Total	\\     \hline 		     
\end{tabular}
\caption{Power counting for $\langle  \mo^{\eta_b}_8 (^1P_1) \rangle$}
\label{tbl:pc}
\end{table}
Numerically, $\langle  \mo^{\eta_b}_8 (^1P_1) \rangle \approx 1 \,{\rm GeV}^{5}$ (hereafter we use $m_b~=~4.7$~GeV).  We see furthermore, why we have truncated Eqs.~\eqref{eq:pseudorate} and~\eqref{eq:scalarrate} at two operators.  Including higher orbital angular momentum states brings further $v$ suppression, as the corresponding four-fermion
operators must have additional covariant derivatives.

Alternatively, while this matrix element has not been measured for bottomonium, a related matrix element in charmonium,
$\langle  \mo^{J/\psi}_8 (^3P_J) \rangle $ has been determined~\cite{Cho:1995ce,Zhang:2009ym}, albeit in linear combination with $\langle  \mo^{J/\psi}_8 (^1S_0) \rangle $.  Assuming the two terms are comparable,
we have $10^{-2}\, {\rm GeV}^5 \approx \langle  \mo^{J/\psi}_8 (^3P_0) \rangle \approx \frac{1}{3} \langle \mo^{\eta_c}_8 (^1P_1) \rangle $.\footnote{ Recently, groups have attempted to determine $\langle  \mo^{J/\psi}_8 (^3P_0) \rangle $ 
\cite{Ma:2010jj,Butenschoen:2011yh,Butenschoen:2011ks} and even $\langle  \mo^{\Upsilon}_8 (^3P_0) \rangle $ \cite{Wang:2012is}, treating QCD at NLO.  This is a highly nontrivial analysis on both the theoretical and experimental sides.  
Unfortunately, these results have 
large uncertainties and currently take negative central values.  Since the $a$ decay rate is proportional to the nonperturbative matrix element, not its mod squared, we cannot use these determinations.  Were these 
negative values to persist, the thought experiment of coupling a light pseudoscalar $a$ to $b$ quarks provides evidence for the breakdown of factorization in NRQCD, recently discussed in \cite{Fleming:2012wy}. The nonperturbative 
contributions to the $a$ decay to bottomonia would have to be positive.  We therefore take our estimates of the long-distance matrix elements from power counting in bottomonium and from the order of magnitude provided by the charmonium data,
which are in rough agreement.}
The second approximate equality follows from spin symmetry, which holds that properties of 
quarkonia that differ only by flipping spins should be equal up to $\mo(v^2)$ corrections from chromomagnetic operators once the ratio of the number of spin states is accounted for (hence the factor of 1/3).  
For charmonium, $v^2 = 0.3$, and we can therefore convert from charmonium to bottomonium by taking 
\beq
\langle \mo^{\eta_b}_8 (^1P_1) \rangle \approx  \frac{m_b^5 v_b^7}{m_c^5 v_c^7}   \langle \mo^{\eta_c}_8 (^1P_1) \rangle \approx 0.3 \, {\rm GeV}^{5}.
\label{ref:scaling}
\eeq
Thus, for our computations below, we will split the difference of our estimates and take $\langle \mo^{\eta_b}_8 (^1P_1) \rangle = 0.5 \, {\rm GeV}^{5}$.  Taking this number (and the same 
procedure for other nonperturbative estimates) and the factorized equation~(\ref{eq:factor}), we can proceed to numerically estimate the decay rate from Eq.~\eqref{eq:pseudorate}, which we plot in Fig.~\ref{fig:totalrates}.

The $\eta_b$ is not the only particle which mixes with the $|b \bar b(^1P_1),{\bf 8}\rangle $ state at the leading orders in $v$. We show the full list of these states in Table~\ref{tbl:1p1}. Although $h_b$ and $^1F_3$ will also have non-vanishing overlap with this states, we will further neglect them in our discussion since technically they overlap at higher order in $v$. Nonetheless, the $^1D_2$ state should have the same overlap with  $|b \bar b(^1P_1),{\bf 8}\rangle $ as $\eta_b$ and 
must be considered as well. It is straightforward to see why this has the same power counting.  Like the $\eta_b$, it has no relative spin-flip relative to $^1P_1$, just a change in orbital angular momentum by one unit.  

\begin{table}[ht]
\begin{tabular}{|l|l|}

\hline
Hadron & $\mo(v)$ \\ \hline
$\eta_b$ & $v$ \\ \hline
$^1D_2$ & $v$ \\ \hline
$\chi_{bJ}$ & $v^{3/2}$ \\ \hline
$h_b$ & $v^2$ \\ \hline
$^1F_3$ & $v^2$ \\ \hline
		     
\end{tabular}
\caption{Hadrons whose Fock state expansion contains a $b \bar b (^1P_1)_8$ component, and the order at which it appears ({\it cf.} Eq.~\eqref{eq:vexp}).}
\label{tbl:1p1}
\end{table}
Similarly, we can understand the $\chi_{bJ}$'s power counting as it has the same orbital angular momentum as $^1P_1$, but differs by spin-flip as it is $^3P_J$, bringing in an extra factor of $v^{1/2}$ in the amplitude.\footnote{The power 
counting of \cite{Bodwin:1994jh} used in Table \ref{tbl:1p1} assumes that $m_Q v \gg \Lambda_{QCD}$.  A more conservative approach takes $m_Q v \sim \Lambda_{QCD}$, as is done in \cite{Fleming:2000ib,Brambilla:2004jw,Brambilla:2006ph}.  These parametrics seem disfavored for bottomonium, which is why do not use them in our analysis, but may relevant for analyzing charmonium, since $m_c v \approx$ 700 MeV.  For completeness though, we mention the effect adopting them would have on our results.
In this regime, one no longer has a hierarchy between electric and magnetic gluons, and so subleading Fock states with a single magnetic gluon are only suppressed by $\mo(v)$ \cite{Brambilla:2006ph}.  This means that each $\chi_{bJ}$ angular 
momentum state would be the same order as $\eta_b$ or $^1D_2$ and collectively they would dominate.  Furthermore, in the $\eta_b$ itself, the $^3S_1$ component of its Fock state ({\it cf.}~Eq.~\ref{eq:vexp}) would come in 
at $\mo(v)$, the same as $^1P_1$. }
Using the results from Table \ref{tbl:1p1}, we get 
\beq
\langle  \mo^{\eta_b}_8 (^1P_1) \rangle \, \approx \, \langle  \mo^{^1D_2}_8 (^1P_1) \rangle \, \approx \, 3 \langle  \mo^{\chi_{bJ}}_8 (^1P_1) \rangle .
\label{eq:me}
\eeq
The 3 in Eq.~\ref{eq:me} comes from the suppression of the $\chi_{bJ}$ rate due to an additional power of $v$.  However, since there are three nearly mass-degenerate $\chi_{bJ}$ states, it is 
effectively leading order as well.  We base our numerical estimates for the other states on the $\eta_b$, with 
$\langle  \mo^{\eta_b}_8 (^1P_1) \rangle$ = 0.5 GeV$^5$.  For the unmeasured $\eta_b$, $\chi_{bJ}$ and all $^1D_2$ masses needed for Fig.~\ref{fig:totalrates} and~\ref{fig:br}, we use the values determined using the potential model in Ref.~\cite{Ebert:2011jc}.\footnote{Practically we take from data only $\eta_b(1)$ resonance mass, however it has been recently claimed in~\cite{Dobbs:2012zn}, that the $\eta_b(2)$ has been also experimentally observed.}
Eqs.~\ref{eq:pseudorate} and \ref{eq:scalarrate} involve an additional nonperturbative term, $\langle  \mo^{{\rm i}}_8 (^3S_1) \rangle$. Of our states of interest, the dominant matrix elements are $\langle \mo^{\chi_{bJ}}_8 (^3S_1)\rangle \approx$ 0.3 GeV$^3$, with $J = 0,1,2$, which we have estimated by power counting.

Before proceeding, we make one final comment on the decay of scalar vs.~pseudoscalar $a$.  Since the pseudoscalar rate has a pole in the $^1P_1$ term for $m_{\rm onia}/m_a = 1$, while the scalar rate vanishes in this limit, the former will be larger.  For $m_a$ = 12 GeV, the ratio
is a factor of 4.  Should we find a spin-0 state in this mass range, improving on our tree-level, estimated rates, especially by an independent measurement of nonperturbative matrix elements, may allow a determination of the P eigenvalue of $a$.

\section{Binding Regime}
\label{sec:binding} 

For mass splittings $\gtrsim \mo(m_b v)$ between our light, spin-0 particle, $a$, and our bottomonium decay product, NRQCD allows us to easily compute the $a$'s decay rates in Eqs.~\eqref{eq:pseudorate} and~\eqref{eq:scalarrate}.  As a local 
effective field theory, we can use the expansions in $\alpha_s,\, m_b,$ and $v$ to systematically capture leading contributions with a handful of operators.  
For smaller mass splittings, the 
calculation becomes more involved.  We argued in Section~\ref{NRQCDreview} that we expect a breakdown when the momentum of the emitted gluon becomes comparable to the momentum transfer within the bound state.  We can see this operationally 
as follows.  In performing the NRQCD calculation, we expanded about the limit of zero relative quark momentum, which gave us a series in $v^2$ associated with higher orbital excitations.  However, as we squeeze the support of the amplitude into a 
region of width $\epsilon \, m_b$, our expansion switches to powers of $v^2/\epsilon$~\cite{Beneke:1997qw}.  Thus, for energy splittings of $\mo(m_b v^2)$, which correspond to a mass splitting of 
$\mo(m_b v)$,
we can only calculate if we can sum up infinite towers of operators built on those giving the leading contributions for larger splittings. 
Furthermore, we see that for even smaller mass splittings ($\epsilon \ll v^2$), we lose calculational control entirely, as suppressed operators become important, since 
$v^n (v^2/\epsilon)^m \gtrsim$ 1.

We do not do so in this work, but the OPE provides a means to sum the infinite towers of operators necessary for $\Delta m \sim \mo(m_b v)$
(for a review of OPEs in $b$-physics, see \cite{Manohar:2000dt},
and an OPE for bottomonium decay is found in~\cite{Rothstein:1997ac}).\footnote{An alternative method for calculation could be to use the effective field theory potential-NRQCD (pNRQCD).  It is designed to operate at mass scales $\lesssim \mo(m_b v)$, where one 
integrates out the gluons responsible for binding, those with $(E,\, p) \,\sim\, (m_b v^2,\, m_b v)$, leading to a spatially non-local, instantaneous potential (see~\cite{Brambilla:1999xf,Pineda:2011dg} for reviews).}
We define
\beq
{\cal T} = i \, \int d^4x \; T \Bigl( \ml_{a \bar b b}(x) \ml_{a \bar b b}(0) \Bigr),
\label{eq:transitionop}
\eeq
where $\ml_{a \bar b b}$ is one of the terms in Eq.~\ref{eq:pscoupling}.  Then, by the optical theorem,
\beq
\Gamma^{a \to b \bar b-{\rm states}\, + X} = \frac{{\rm Im} \langle a| {\cal T} | a \rangle}{m_a}.
\label{eq:optical}
\eeq
Performing the OPE on ${\cal T}$ gives us an infinite series of NRQCD operators 
whose expectation values we can sum into 
nonperturbative structure functions, as in~\cite{Beneke:1997qw,Rothstein:1997ac}. 

Although we do not perform the OPE calculation (which is expected to give a rather precise answer for $\Delta m \sim \mo(m_b v)$), we roughly estimate the 
decay rate to bottomonia using a different technique. We notice that 
when $m_a$ becomes comparable to or smaller than the heaviest $\eta_b$ mass (= 11 GeV), certain decay rates will receive significant effects from $a-\eta_b$ mixing.  These are
\begin{enumerate}
 \item $a \to gg$
 \item $a \to \eta_b + X$
 \item  $a \to$ (open-flavor) $b \bar b$
\end{enumerate}
 We will spend the rest of the section discussing 1) and 2), while leaving 3) for an 
Appendix (since we use the approach of \cite{Drees:1989du}, albeit with updated data).  After we compute the mixed content of physical states, the $a \to gg$ rate
will follow readily.  We compute rate $a \to \eta_b + X$ solely due to hadronic transitions between bottomonia after mixing with the $a$.  This will give us an 
order of magnitude estimate on the rate for $\Delta m \leq \mo(m_b v)$, where the calculation is either involved or very difficult.  Since we are in a region where 
parametric control is breaking down, it is difficult to quantify how much of the total rate these mixing effects represent.  However, at those locations where the $a$ is 
degenerate with an $\eta_b$ and mixing is maximal, we should expect the decay rate due to mixing to dominate, and thus these points give us an 
approximate upper bound.

\subsection{$a-\eta_b$ Mixing and $gg$ annihilation}
\label{subsec:mixing}

\begin{figure}[ht!]
\centering
\includegraphics[width=0.5\textwidth]{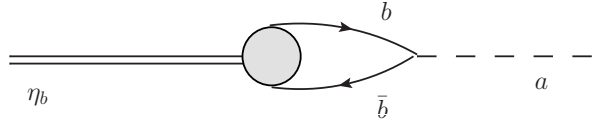} 
\caption{Feynman diagram for $a-\eta_b$ mixing.  The blob represents the nonperturbative contribution of the radial wavefunction at the origin, 
$R_{\eta_b (n)}(0)$.}
\label{fig:mix}
\end{figure}

Since the $a$ has the same quantum numbers as the $\eta_b$s, it will mix them.\footnote{Since parity is broken in the electroweak sector, the $a$ will mix the the parity-
even $\chi_{b0}$ as well, but this rate is suppressed by weak coupling, a loop factor, and 
the $W$ mass so we neglect it.} We can calculate the mixing from the diagram in Fig.~\ref{fig:mix}.  
Quantifying the conversion of a $^1S_0$ $b \bar b$ singlet into
an $\eta_b$ hadron does not require the full machinery of NRQCD, but can be described simply via the hadronic radial wavefunction at the origin \cite{Drees:1989du}.\footnote{In this section, we switch to working with a Schr\"{o}dinger potential model (see {\it e.g.}~\cite{Buchmuller:1981aj}).  The mixing, annihilation, and hadronic transition rates we need can be written very easily in terms of the $b \bar b$ wavefunction, and the numerical value extracted from experiment.  
One can recast the wavefunction in terms of NRQCD matrix elements, {\it e.g.}~for the $\eta_b$ radial wavefunction, 
$R_{\eta_b} ({\bf r}) = \sqrt{\frac{2 \pi}{N_c}} \langle 0| \chi^\dag(-{\bf r}/2) \psi ({\bf r}/2) | \eta_b \rangle $ \cite{Bodwin:1994jh}.}  The mass matrix that controls $a-\eta_b$ mixing thus contains the following off-diagonal entries:
\beq
\delta m_{a-\eta_b(n)}^2 = y_b \sqrt{\frac{3}{4 \pi} m_{\eta_b(n)}} \, |R_{\eta_b(n)} (0)|,
\label{eq:aetamixing}
\eeq
One can obtain the nonperturbative quantity $|R_{\eta_b} (0)|$ from experiment.  
While there is little data for the $\eta_b$s directly, by spin symmetry we expect the
wavefunction values of the $\Upsilon$ to be the same up to suppressed chromomagnetic effects.  We can therefore get a numerical value for Eq.~\ref{eq:aetamixing} 
from the PDG~\cite{Nakamura:2010zzi} via the following formula:
\beq
|R_{\Upsilon(n)} (0)|^2 = \Gamma( \Upsilon (n) \rightarrow e^+ e^-) \frac{9 \, m^2_{\Upsilon(n)}}{4 \alpha^2} \left( 1 - \frac{16 \, \alpha_s(m_\Upsilon)}{3 \pi} \right)^{-1}.
\label{eq:wfxndecay}
\eeq
Eqs.~\eqref{eq:aetamixing} and~\eqref{eq:wfxndecay} give us the mixing matrix,
\beq
M^2 = \left( \begin{array}{cccc}
m_a^2 - i m_a \, \Gamma_a & \delta m_{a - \eta_b(1)}^2 & \ldots & \delta m_{a - \eta_b(6)}^2 \\
\delta m_{a - \eta_b(1)}^2  & m_{\eta_b(1)}^2 - i m_{\eta_b(1)} \, \Gamma_{\eta_b(1)} &  \ldots & 0 \\
\vdots & 0 & \ddots & 0 \\
\delta m_{a - \eta_b(6)}^2 & 0 & 0 & m_{\eta_b(6)}^2 - i m_{\eta_b(6)} \, \Gamma_{\eta_b(6)}
\end{array} \right).
\label{eq:matrix}
\eeq
We include six $\eta_b$ states, corresponding to the six observed $\Upsilon$ particles.  Other than the $\eta_b(1,\,2)$, their masses are unmeasured and have to be taken from calculation (Refs.~\cite{Barbieri:1975ki, Barbieri:1981gj, Buchmuller:1981aj, Gupta:1989jd,Ebert:2011jc}).  We give their values, along with
$|R_{\eta_b(n)}(0)|^2$ in Table~\ref{tbl:eta}.  
\begin{table}[ht]
\begin{tabular}{|l|c|c|}

\hline
$\eta_b (n)$ & $m_{\eta_b (n)}$ (GeV) & $|R_{\eta_b(n)}(0)|$ (GeV)$^{3/2}$ \\ \hline
1 & 9.4 & 2.7 \\ \hline
2 & 10.0 & 1.9 \\ \hline
3 & 10.3 & 1.6 \\ \hline
4 & 10.6 & 4.7 \\ \hline
5 & 10.85 & 4.7 \\ \hline
6 & 11.0 & 3.0 \\ \hline
		     
\end{tabular}
\caption{Masses and wavefunctions at the origin for the $\eta_b$ states.}
\label{tbl:eta}
\end{table}
The various nonzero widths, $\Gamma$, give us
a non-Hermitian matrix, but it is still symmetric and diagonalizable via a complex rotation among the $a$ and $\eta_b(n)$.  For the premixed widths, we take $\Gamma_a$ to include decays to $c \bar c$, $\tau^+ \tau^-$, $gg$, and in the regime 
of validity for NRQCD ($\gtrsim$ 11.5 GeV), the bottomonium + $X$ decays determined in Section~\ref{sec:nrqcd}.  
For the decay rates of the $\eta_b(n)$ themselves, we consider three classes of decays: annihilation to $gg$, hadronic transitions to lower lying $\eta_b$, 
and for the $\eta_b(5)$ and $\eta_b(6)$, decays to $B$ mesons.\footnote{Unlike 
the $\Upsilon$ system, where the $4S$ famously decays to two $B$ mesons, the $\eta_b(4S)$ is disallowed by CP and P symmetry from decaying to $B \bar B$, and 
kinematically forbidden to go to higher mass, higher spin mesons.}
Once again, our approximation is to neglect any decay of the $a$ to bottomonium that does not involve mixing for $m_a \lesssim$ 11 GeV.  

Computing the decay to gluons is straightforward.  
The $\eta_b$'s rate is
\beq
\Gamma(\eta_b(n) \rightarrow gg) = \frac{\alpha_s(m_{\eta_b(n)})^2}{3\, m_{\eta_b(n)}^2} |R_{\eta_b(n)} (0)|^2.
\label{eq:gludecay}
\eeq
We also have the $a$'s ``own decay'' $a \rightarrow gg$.  
For the one-loop contributions to this rate, we use the formulas of Refs.~\cite{Drees:1989du, Bellazzini:2009xt}, including $t \bar t,\, b \bar b$, and $c \bar c$ loops.  Diagonalizing the matrix in Eq.~\eqref{eq:matrix} and rotating to mass eigenstate basis, we get for our physical $a$ particle
\beq
a_{\rm phys.} = c_a a + \sum_i c_i \, \eta_b(i),
\eeq
where the $c_i$ are generically complex, and we can get nontrivial interference effects between different decay channels, since these add at the amplitude level,
\beq
\mathcal{M}(a_{\rm phys.} \rightarrow gg ) = c_a \mathcal{M}(a \rightarrow gg) + \sum_i c_i \, \mathcal{M}(\eta_b(i) \rightarrow gg).
\label{eq:superpos}
\eeq
In fact, the changes to masses of the physical eigenstates from the naive unmixed values are negligible, but we see the change to $\Gamma(a \rightarrow gg)$ is pronounced ({\it cf.}~Fig.~\ref{fig:totalrates}).
However, compared to the other decay rates of the $a$, the branching ratio to gluons is still small.  For particular models though, if the effective coupling to $\tau^+ \tau^-$ gets extra suppressions beyond the Yukawa coupling, this rate could be important and is used in the buried Higgs \cite{Bellazzini:2009xt}.

\subsection{Bottomonium hadronic transitions}
\label{subsec:mp}

In this subsection we estimate decays of the $a$ to bottomonia via its mixing with the $\eta_b$ states (Eq.~\ref{eq:aetamixing}), which undergo a subsequent 
hadronic transition.  In the case of decays into gluons we could easily calculate the $a$'s ``native'' decay for $m_a \in [9.4, 11]$ GeV.
When the decays proceed into bottomonia this computation involves an infinite number of operators, as discussed at 
the beginning of this Section.   
We will therefore estimate the $a$'s decay rates to bottomonia in this region from the mixing contribution alone.  
If we were in possession of exhaustive data cataloging the decays of all six $\eta_b$s, then this would involve little beyond diagonalizing the mass mixing matrix (Eq.~\ref{eq:matrix}).  
Since only two $\eta_b$s have been observed and their decay channels barely determined, we will need an alternative means to estimate their decay rates, which we will plug into the formula,
\beq
\Gamma(a \rightarrow {\rm bottomonium} + X) =  \sum_i |c_i|^2 \, \Gamma(\eta_b(i) \rightarrow {\rm bottomonium} + X) \theta(m_a - m_{\eta_b(i)}).
\label{eq:superposht}
\eeq
In addition to the approximation of dropping $a$ decays that do not involve mixing, we also sum the rates with identical final states incoherently as the $a$ only mixes strongly with one bottomonium at a time,
and we ignore the kinematic difference between $m_a$ and $m_{\eta_b(i)}$ since in the mass range where mixing is strong, this will be a small effect.    
The $\theta$-function though, enforces that kinematically forbidden processes cannot occur.  

Although we do not have data on the exclusive channels of $\eta_b$, their spin-flipped partners, the $\Upsilon$s, are much better observed.  Thus, we will use their observed hadronic transition rates.  For the low-lying 
($n=1-3$) states, we can justify using $\Upsilon$s in place of $\eta_b$s by the QCD multipole expansion (for details see \cite{Grinstein:1996gm,Kuang:2006me} and references). This relates transition rates to the 
wavefunctions calculated in a Schr\"odinger potential model, and has had quantitative success in predicting decay rates~\cite{Voloshin:1978hc,Bhanot:1979af,Peskin:1979va,Bhanot:1979vb,Voloshin:1980zf,Yan:1980uh}.
We will use the same approximation as in Section \ref{subsec:mixing}, 
that the $\eta_b$ and $\Upsilon$ wavefunctions will agree up to $\mo(v^2)$, spin-suppressed effects.  Thus, we can take the $\Upsilon$ transition data, which we give in Table \ref{tbl:transit}.\footnote{For the low-lying states where the multipole expansion is more trustworthy, we can 
calculate the self-coupling of the $\eta_b({\rm i})$s to pions, $\sim \eta_b({\rm i}) \eta_b({\rm i}) \pi \pi$.  Using the parametrization of \cite{Grinstein:1996gm} to go between gluons and pions, and the potential model of 
\cite{Peskin:1979va}, we find, for example, the rate of $a \to \eta_b(1) \pi \pi$ through the $\eta_b(1)$ self-coupling to pions $\sim 10^{-8}$ GeV, an order of magnitude below any other rate we calculate, and far below the leading processes.  We thus neglect 
these self-coupling mediated decays.}

For the higher $\Upsilon$ excitations, the multipole expansion fails badly, by as much as two orders of magnitude \cite{Meng:2007tk,Meng:2008bq}.  
Interestingly, the hadronic transition rates for the higher states, are relatively large, $\mo(1)$ MeV, agreeing with what we find in the nearby perturbative NRQCD region ({\it cf.}~Fig.~\ref{fig:totalrates}).  
It has been suggested that one must
take into account the effect on-shell $B$-meson states can have on transition rates.  
Attempting to parametrize these, the authors of Ref.~\cite{Meng:2007tk,Meng:2008bq} could predict rates at the factor-of-two level.
More recently, data on the hadronic transitions of $\Upsilon(5)$ have shown the presence of nearby narrow resonances responsible for the enhancement of its decay rate \cite{Collaboration:2011gja, Belle:2011aa, Adachi:2012im}.  
The transition proceeds in two steps, 
$\Upsilon(5) \rightarrow Z_b \pi$ followed by $Z_b \rightarrow \Upsilon(n) \pi$ ($n=1-3$), where $Z_b$ is the new, charged, spin-1 particle, likely a molecular bound state of $B$ mesons.  By heavy quark spin-symmetry, an analogous molecular state
with spin-0 is predicted to exist, $W_{b0}$, with comparable couplings to the $\eta_b$ as $Z_b$ has to the $\Upsilon$ \cite{Bondar:2011ev, Voloshin:2011qa, Mehen:2011yh}.  Thus, we should expect the same enhanced decay rates for the pseudoscalar bottomonia as for the vectors.
This justifies using the $\Upsilon$ hadronic transition data for the higher excitations (which we present in Table \ref{tbl:transit}) in Eq.~\ref{eq:superposht} for the $\eta_b$ decays.\footnote{Our use of data breaks down for the $\Upsilon(6)$, 
whose dipion transitions to other $\Upsilon$ states are unmeasured.  Thus, for its rates we use a combination of those provided in 
\cite{Meng:2007tk} and an assumption that ${\rm BR}[\Upsilon(5) \rightarrow \Upsilon(5-n)] \approx {\rm BR}[\Upsilon(6) \rightarrow \Upsilon(6-n)]$ for the others.}
In detail, due to spin and kinematic factors there may be $\mo(1)$ differences in the $\eta_b$ and $\Upsilon$ rates, but for the order of magnitude estimate we are trying to get for $a$ decays in the binding regime, these are beyond the order to which we are working.

%If we assume the interactions with $B$-mesons are important for 
%the size of the higher-state hadronic transitions, as in \cite{Meng:2007tk,Meng:2008bq}, then the results we show in the Appendix, that the decay rate of $\eta_b$ to $B$-mesons is comparable to that of $\Upsilon$ to $B$-mesons 
%({\it cf.~}Eq.~\ref{eq:oniaB}), again justifies using the $\Upsilon$ hadronic transition data (which we present in Table \ref{tbl:transit}) in Eq.~\ref{eq:superposht} for the $\eta_b$ decays.

For the higher states, there can also be sufficient phase space to decay through heavier pseudo-Goldstones such as $\eta$ and $K$.  We list these additional processes in Table \ref{tbl:transit}, which can have comparable rates to the dipion modes, but 
we do not include them in our branching ratio calculations since we cannot do so systematically, lacking data for the $\Upsilon(5,\,6)$ data for the single-$\eta$ and/or $KK$  channels, and we are only attempting an order of magnitude estimate.  
\begin{table}[t]
\begin{tabular}{|c|c|c|c|}

\hline

Initial state i= & $\sum_j \Gamma(\Upsilon({\rm i}) \to \Upsilon({\rm j}) \pi \pi)$ MeV & Multipole valid & Other important transitions \\ \hline
2 & 9 $\times 10^{-3}$& Yes & - \\ \hline
3 & 2 $\times 10^{-3}$ & Yes & $\Upsilon(2) \gamma \gamma$ \\ \hline
4 & 5 $\times 10^{-3}$ & No & $\Upsilon(1) \eta$ \\ \hline
5 & 1.5 &  No & $\Upsilon(1) KK,\,\Upsilon(1,\,2) \eta$?  \\ \hline
6  & 3 & No & $\Upsilon(1,\,2) KK?,\, \Upsilon(1-3) \eta$? \\ \hline

\end{tabular}
\caption{Total hadronic transition rates for the dipion decays, $\Upsilon({\rm i}) \to \Upsilon({\rm j}) \pi \pi$, which we use for the $\eta_b$ transition rates in Eq.~\ref{eq:superposht}.}
\label{tbl:transit}
\end{table}

In the end, we find that the rate of $a \to \eta_b(n) \pi \pi$ from hadronic transition after mixing
is significant only where it has large overlap with the $\eta_b(5)$ and $\eta_b(6)$ states ({\it cf.~}Figs.~\ref{fig:totalrates} and \ref{fig:br}).

\section{Discussion}
\label{sec:discussion}

\subsection{Rate Plots}
\label{subsec:plots}

We present our results for the $a$'s decay rates and branching ratios for the region 9-15 GeV.  For these figures, we restrict to the case of pseudoscalar $a$.
Most simple to describe are those to open fermion pairs, 
partonic-$b \bar b,\, c \bar c$, and $\tau^+ \tau^-$.  At tree level, this rate is given by
\beq
\Gamma(a \rightarrow f \bar f) \,=\, N_c \frac{G_F}{2\sqrt{2} \pi} m_f^2 \, m_a \left( 1 - \frac{4 m_f^2}{m_a^2}  \right)^{1/2}.
\label{eq:fermiondecay}
\eeq
For the partonic calculation of quark rates, we also apply the NLO QCD correction given in Ref.~\cite{Drees:1989du}.  For $a \to gg$, we use the result of 
Section \ref{subsec:mixing} where took into account the effects of $a-\eta_b$ mixing.
Mixing also affects the turnover to production of $b$-flavored mesons, where we have used a phenomenological interpolation ({\it cf.} Eq.~\ref{eq:interp}) to go between a partonic QCD calculation for $b$-quarks with $m_b = m_B$ (Eq.~\ref{eq:fermiondecay}) and the hadronic mixing formula, which uses Eqs.~\ref{eq:bmodel} and \ref{eq:mixedcoeffs}.  
As discussed at the beginning of Section \ref{sec:binding} and in Section \ref{subsec:mp}, for $m_a \lesssim$ 11 GeV, we estimate its decay rate to bottomonia only through its mixing with $\eta_b$, which undergo hadronic transition decays to give $a \to \eta_b \, \pi \pi$.  We use the $\Upsilon$ transition rates as 
approximately those of the $\eta_b$ by spin symmetry in Eq.~\ref{eq:superposht}.  
Lastly, we have the process $a \to {\rm bottomonium}\, + X$ calculated systematically in Sec.~\ref{sec:calcNRQCD} for $m_a \gtrsim$ 11 GeV, using Eq.~\ref{eq:factor}.  
We see that the hadronic transition after mixing and NRQCD computations come within an order of magnitude where we switch from 
one method to the other.  This gives us reason to believe a detailed study could find a smooth interpolation of the total $a \rightarrow {\rm onia}\, + X$ rate as
one passes from the regime of $\Delta m_{a-{\rm onia}} > m_b v$ to $\Delta m_{a-{\rm onia}} \sim m_b v$ at values roughly comparable to those we depict.

\begin{figure}[t]
\centering
\includegraphics[width=0.9\textwidth]{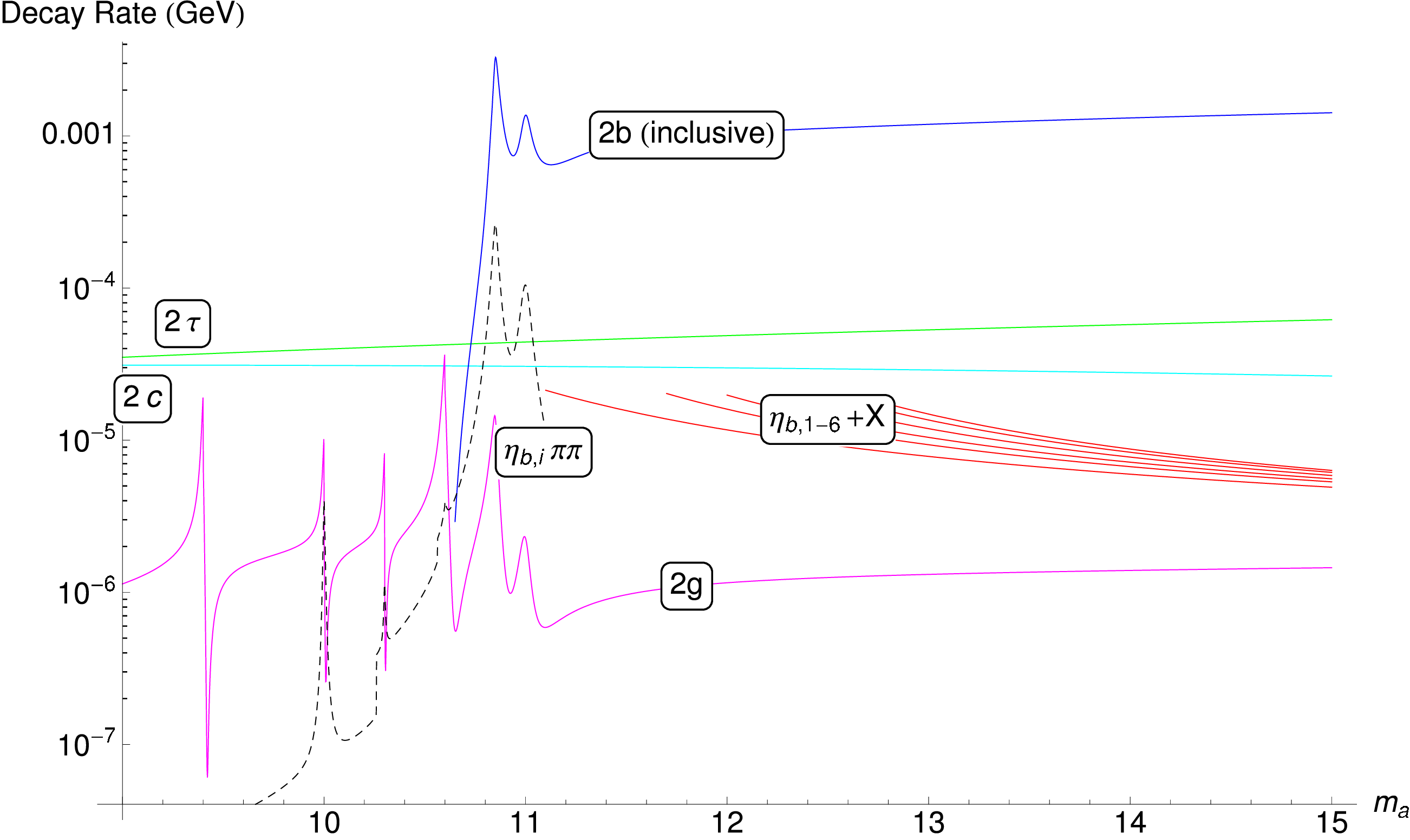} 
\caption{The total decay rates for pseudoscalar $a$ for each channel in GeV as a function of $m_a$.  
The $b \bar b$ rate is taken from the interpolation equation (\ref{eq:interp}).  For visual simplicity, we have only included the $\eta_b$.  We plot the hadronic transition rates from $a-\eta_b$ mixing
up to a mass of 11.1 GeV, at which point we switch to our NRQCD calculation, Eq.~\ref{eq:pseudorate}, for the decay $a \rightarrow \eta_b(1)\,+X$.  We do not use NRQCD for $m_a - m_{\eta_b(n)} <$  1.7 GeV for any $n$. 
The $\eta_{b,\,i} \pi \pi$ curve is dashed as we have only included those contributions from $a-\eta_b$ mixing.  It is thus meant as an order of magnitude approximation, rather than a systematic calculation.}
\label{fig:totalrates}
\end{figure}

\begin{figure}[t]
\centering
\includegraphics[width=0.9\textwidth]{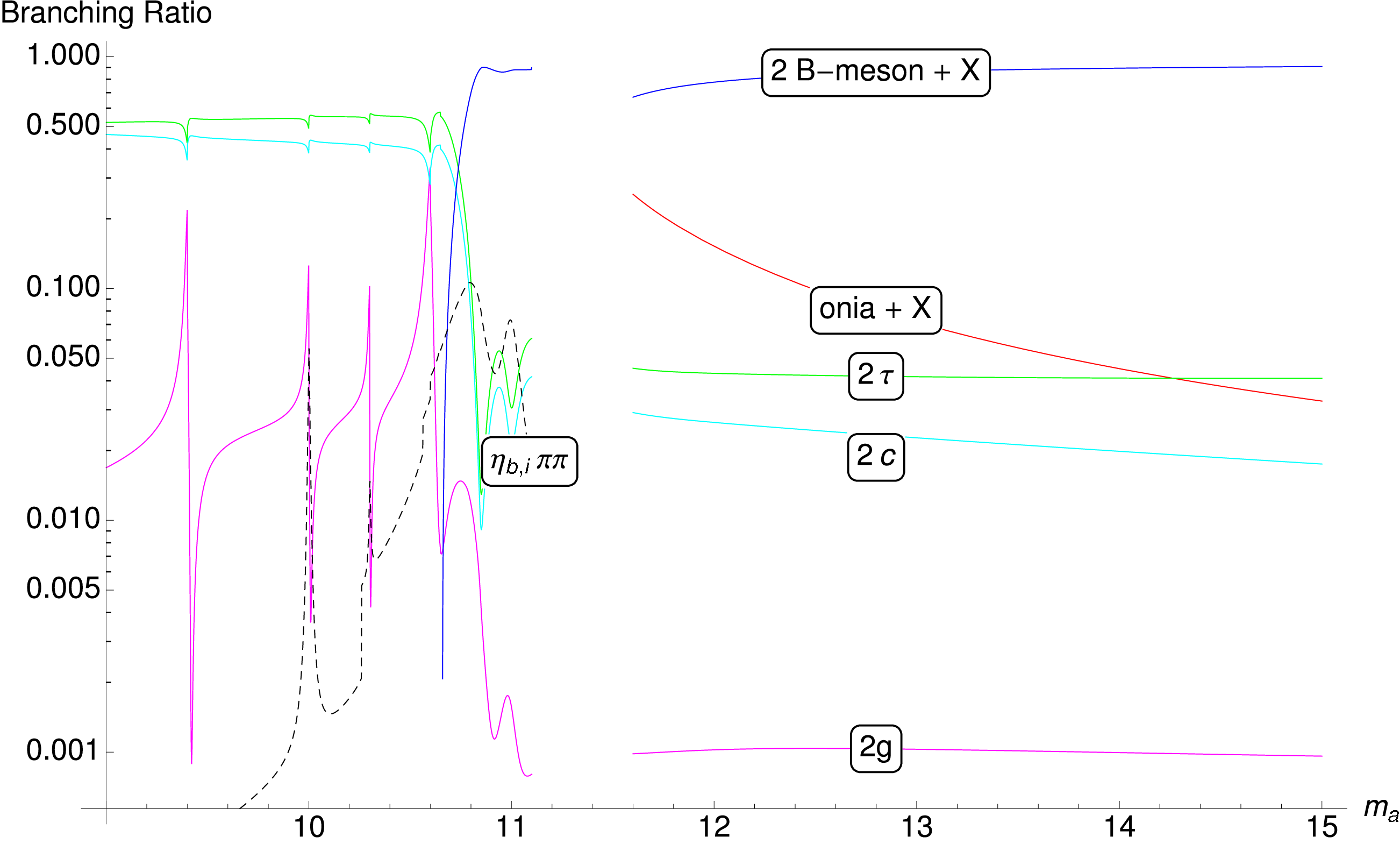} 
\caption{Branching ratios to the various decay products of an $a$ pseudoscalar.     
We have summed over the four $\eta_b$, nine $\chi_{b\, J}$, and two $^1D_2$ sub-threshold states to get an overall $a \rightarrow {\rm onia}\, + X$.  We
cease to consider a bottomonium decay of the $a$ in NRQCD below $m_a$ = 11.6 GeV.  We show the dominant contribution for smaller $\Delta m$ though, 
from the decays $a \rightarrow \eta_b \pi \pi$, starting at $m_a$ = 11.1 GeV, where we begin to see mixing effects with the heaviest $\eta_b$.  
The $\eta_{b,\,i} \pi \pi$ curve is dashed as we have only included those contributions from mixing.  It is thus meant as an order of magnitude approximation. For the blue curve, we have subtracted the 
contributions of sub-threshold bottomonia from the inclusive $b \bar b\, + X$ rate given by Eq.~\ref{eq:interp} to give the rate to open-flavored mesons.}
\label{fig:br}
\end{figure}

In Fig.~\ref{fig:totalrates}, for visual simplicity, we only show the $\eta_b$ NRQCD-derived decay rates, even though $^1D_2$ and $\chi_{bJ}$ are large as well.
For each of the six states, we cut off the curve when the $a-\eta_b$ mass splitting becomes smaller than $\mo(m_b v)$.  To plot branching ratios in Fig.~\ref{fig:br}, 
we need a full accounting of the various decay rates, and thus include all of our leading NRQCD-derived rates.  Since the $\eta_b(5,\,6)$ decay overwhelmingly to 
$B$-mesons, which experimentally look like $b \bar b$ decays, we do not add them, nor any of the other above-threshold bottomonia, to the ``onia + $X$'' curve.
The NLO QCD computation we use for the decay to $b \bar b$ is really an inclusive rate for all hadronic channels that contain a $b \bar b$ pair and a perturbative 
gluon.  This therefore includes the $a \to {\rm bottomonium}\, + X$ final state, and so for Fig.~\ref{fig:br}, we subtract our rate obtained from Eq.~\ref{eq:pseudorate}
for the sub-threshold bottomonia from the inclusive one (Eq.~\ref{eq:interp}) to get branching ratios for open-flavor channels versus those that cannot give two 
$B$-mesons.

For the sub-threshold bottomonia, we use the NRQCD calculation down to a mass of 11.6 GeV, as this leaves a large enough splitting to the heaviest such state, 
$\eta_b(4)$, that $X$ can involve many possible decay channels.  We have left the range [11.1, 11.6] GeV blank in Fig.~\ref{fig:br}, as we lack a compelling way to estimate our way 
through it.  Since the mass splitting in this range is $\mo(m_b v)$, it is possible in principle to use an OPE based in Eq.~\ref{eq:transitionop} to compute the rate, provided one can
numerically determine the resulting nonperturbative structure function.  It would be an interesting exercise to see how extending the NRQCD results down to this mass splitting 
compares with the black curve we have estimated from $a-\eta_b$ mixing followed by hadronic transition.


\subsection{Observability}
\label{sec:disc}

Looking at Fig.~\ref{fig:br} we see that the bottomonium channel is almost always subdominant, either to $b \bar b$ or to $\tau^+ \tau^-$ and $c \bar c$. However, in some parts of parameter space it is not negligible and can account for almost 30\% of the entire branching ratio. Bottomonium does not look very different from a regular QCD jet (even though it typically has fewer tracks) and therefore one can view the enhancement of the bottomonium decay mode as an effective enhancement of the $gg$ channel. Indeed, in the $a$ rest frame, $\eta_b$ is recoiling against a gluon jet, while $\eta_b$ is probably impossible to distinguish from a gluon jet on event-by-event basis. Practically, $b$-tagging does not work on the hadronic decays of sub-threshold quarkonia as they annihilate rapidly to gluons ({\it cf.}~\ref{eq:gludecay}), 
and this is an emergence of ``buried Higgs''-type signatures~\cite{Bellazzini:2009xt}, but at masses far above the ditau threshold.

 Without further evidence for new physics in this particular window, it is perhaps premature to launch a dedicated collider study into the possibility of observing $a \rightarrow {\rm bottomonium}\, + X$ decays.  However, there is still a benefit at this stage in a 
 {\it qualitative} discussion of  how this scenario would appear at the LHC, and how we might go about observing it.  If we take the Higgs mass, $m_h$, to be around 125 GeV and $m_a = 12$ GeV, then even in the Higgs rest frame the decay products 
of the pseudoscalar will be collimated, having $\Delta R \sim$ 0.3.  For the region where $a \rightarrow {\rm bottomonium}\, + X$ decays are important, the gluon is fairly soft, having $p<$ 5 GeV in the $a$ rest-frame.

A standard opening move for analyzing a jetty Higgs is to look at channels where the Higgs itself is boosted, coming from production processes such as $V + h$ and $t \bar t h$~\cite{Butterworth:2008iy}.  
One then attempts to build one large ``Higgs jet'' either through fat 
cones \cite{Falkowski:2010hi} or the Cambridge-Aachen algorithm~\cite{Kaplan:2011vf}.  Having this jet, and after performing some cleaning on it to remove pileup and underlying event \cite{Butterworth:2008iy, Ellis:2009me, Plehn:2009rk},
one can then look for structure consistent with having two dominant clusters of energy with approximately equal masses.  For our case, just as in Ref.~\cite{Falkowski:2010hi}, the particle associated with the subjet, the $a$, is less massive relative to its 
$p_T$ than a QCD jet.   Additionally, as in Ref.~\cite{Falkowski:2010hi}, we can exploit the absence of colored states until the $a$ decays, which gives a different radiation pattern from prompt dijet production.

Despite the similarities in technique appropriate to our scenario and~\cite{Falkowski:2010hi}, and the similarity in final states,  
we have a couple additional
handles.  Firstly, in addition to having the subjets at comparable light masses, their masses are near those of known states, $\eta_b$, $\chi_{bJ}$, and $^1D_2$.  Few of these have been measured directly, but calculations exist for the others 
({\it cf.}~\cite{Barbieri:1975ki, Barbieri:1981gj, Buchmuller:1981aj, Gupta:1989jd, Ebert:2011jc}) likely to be valid to within 1 GeV, which puts them well within experimental resolution.
Additionally, we can make use of the $\eta_b$'s decay into a small number of hadrons.  While the dominant decay products of the $\eta_b$ have not been measured, the $\eta_c$, which similarly annihilates into gluon pairs, typically goes to just 2-4 particles.  
Thus, counting the number of tracks in the decay, and looking for a small number relative to a QCD jet with similar $p_T$ could provide useful additional information.  Should it become necessary, one could even attempt to study in more detail 
the decay properties of these bottomonia states by combining NRQCD with the lattice.

Lastly, we note that for many different arrangements of the $a$'s coupling to SM fermions, over much of the $m_a$ range where bottomonium decays can be
important, there are other, experimentally distinctive decay channels, one can observe.  Below the open $b$ threshold, one could search for events with 
$h \rightarrow (\eta_b + X) (\tau^+ \tau^-)$.  Above it, one could use $b$-tagging on an $a$ that decays to $B$ mesons, and look for the other to decay to a light, sparsely populated subjet.  Thus, despite putting the Higgs in the realm of jet physics, 
one has many handles for digging out the structure inherent to a decay chain passing through quarkonia.  As a final aside, we mention a side benefit of having a scenario such as this realized in nature.  Despite many decades of experimental 
$B$-physics, the $\eta_b$ states remain very poorly studied despite being kinematically the simplest (spinless and $S$-wave).  Nature would have a wry sense of humor to deliver us these particles, and the other previously-unobserved mesons, 
out of its mechanism for electroweak symmetry breaking.

\acknowledgments{We are grateful to Eric Braaten, Zohar Komargodski, Adam Leibovich, Markus Luty, Thomas Mehen, Kirill Melnikov, Antonio Pineda, Ira Rothstein, Iain Stewart, Matthew Strassler, and Raman Sundrum for useful discussions. MB is supported by NSF
grant nsf-phy/0910467 and AK is supported by NSF grant PHY-0855591. The authors thank the Aspen Center of Physics where part of this work was done.}

\appendix

\section{Open Flavor Decays}
\label{subsec:bbbar}

The decay to open $b$-flavored hadrons will dominate for sufficiently large $m_a$.  
One must still determine though, where exactly in the $m_a$ range this process takes over.  This is a somewhat involved question as it brings in issues of quark-hadron duality, IR divergences, and large threshold logarithms;  
we do not attempt to answer it in full.  It is nonetheless interesting from the point of view of decays to bottomonia, as these can be an important subleading effect above the open-flavor threshold, and are kinematically allowed 
even where $B$ meson decays are forbidden.
The zeroth order approximation for $a \rightarrow b \bar b$ is to simply use the partonic formula with $m_b = m_B$, where $B$ is the lightest 
$B$ meson.  Following Ref.~\cite{Drees:1989du}, we wish to improve on this naive treatment by incorporating effects due to $a-\eta_b$ mixing,
discussed in Section \ref{subsec:mixing} in the threshold region for $B$ meson production.  We know the partonic calculation is making a kinematic mistake in the threshold region, and this is our attempt to partially correct it.  
The pseudoscalar decay into $b$ quarks is an $S$-wave process by parity.  However, the decay into the lightest allowed physical mesons, 
$a \rightarrow B^* \bar B \,+\, B \bar{B}^*$ is $P$-wave.  Thus, we include the modification of the threshold 
rate due to the presence of $\eta_b(5)$ and $\eta_b(6)$, with their appropriate $P$-wave coupling to $B$-mesons.

One important qualitative difference between $S$ and $P$-wave is the momentum dependence near threshold.  The physical, $P$-wave, decay will be proportional to the decay products' momenta, which 
goes to zero at threshold, while the partonic $S$-wave channel receives no such suppression.
We write out the decays of $\eta_b$ and $\Upsilon$ to the lightest $B$ mesons so that we might use data from the latter to estimate the former,
\beqa
\label{eq:bmodel}
\mathcal{M}(\eta_b \rightarrow B^* \bar B) &=& a_B\, k \delta_{\lambda 0} \, d^0_{0 \lambda} \label{eq:bmesonrates} \\  \nonumber
\mathcal{M}(\eta_b \rightarrow B^* \bar{B}^*) &=& a_B\, k \delta_{\lambda \bar \lambda} \, d^0_{0 \lambda_f} \\ \nonumber
\mathcal{M}(\Upsilon \rightarrow B \bar{B}) &=& a_B\, k d^1_{\lambda_i 0} \\ \nonumber
\mathcal{M}(\Upsilon \rightarrow B^* \bar B) &=& a_B \, k \lambda_f d^1_{\lambda_i \lambda_f} \\ \nonumber
\mathcal{M}(\Upsilon \rightarrow B^* \bar{B}^*) &=& a_B\, k d^1_{\lambda_i \lambda_f}, 
\eeqa
and similarly for $B \rightarrow B_s$.  The helicities are $\lambda_i$ for the decaying meson, $\lambda,\, \bar \lambda$ for the final state particles, and $\lambda_f = \lambda - \bar \lambda$.  The norm of the three-momentum of the 
decay products in the $\eta_b$ or $\Upsilon$ rest frame is $k$.  This factor arises from the $P$-wave nature of the decays.  
To avoid introducing extra notation, we have used the same symbol throughout, but $k$ will differ with each process.  The terms $d^J_{\lambda \mu}$ are Wigner's little-$d$ functions.
The overall factor $a_B$ is a three-meson coupling and is assumed to be the same for all of Eq.~\ref{eq:bmesonrates} since each line can be obtained from another via spin flips.  

In practice, the assumption that these different 
channels are all controlled by a single coupling holds only to within  factors of two in the $B$ decays of the $\Upsilon$, and at the order of magnitude level for $B_s$.  We nonetheless make it as we can only hope to understand threshold 
effects up to $O(1)$ factors anyway.  Additionally, the branching ratio of $\Upsilon(5S)$ to $B$ mesons is nearly three times
that to $B_s$, so we accept a larger uncertainty in the subdominant process.  We fix the couplings $a_B$ and $a_{B_s}$ such that the overall rate of $\Upsilon(5)$ into the three channels of Eq.~\ref{eq:bmodel} agrees with data.
The inclusion of $B_s$ decays is a change we have made from \cite{Drees:1989du}, as experimental evidence for such processes in $\Upsilon(5S)$
decays was just starting to be reported at the time.  Calculating the ratio of rates from Eq.~\ref{eq:bmesonrates}, we get
\beqa
\label{eq:oniaB}
\Gamma(\eta_b(5) \rightarrow B \;{\rm mesons}) &\approx& 0.9 \, \Gamma(\Upsilon(5) \rightarrow B \;{\rm mesons}) \\ \nonumber 
\Gamma(\eta_b(5) \rightarrow B_s \;{\rm mesons}) &\approx& 0.65 \, \Gamma(\Upsilon(5) \rightarrow B_s \;{\rm mesons}) \\ \nonumber 
\Gamma(\eta_b(6) \rightarrow B \;{\rm mesons}) &\approx& \Gamma(\Upsilon(6) \rightarrow B \;{\rm mesons}) \\ \nonumber 
\Gamma(\eta_b(6) \rightarrow B_s \;{\rm mesons}) &\approx& \Gamma(\Upsilon(6) \rightarrow B_s \;{\rm mesons}),  
\eeqa
where we have summed the first two lines of Eq.~\ref{eq:bmesonrates} for the $\eta_b$ rates and the last three lines for $\Upsilon$.  The $\Upsilon(5)$ is a well-studied particle, and so we can use the measured decay rates to estimate $a_B$ and 
$a_{B_s}$.  While the total width of the $\Upsilon(6)$ is known to within 20\%, the individual hadronic channels of interest are currently unknown.  Therefore, we assume its branching ratios to the various $B$ and $B_s$ final states are the same
as those of the $\Upsilon(5)$, which should not induce a significant further error in our admittedly approximate treatment of the threshold region.  
\begin{figure}[ht!]
\centering
\includegraphics[width=0.6\textwidth]{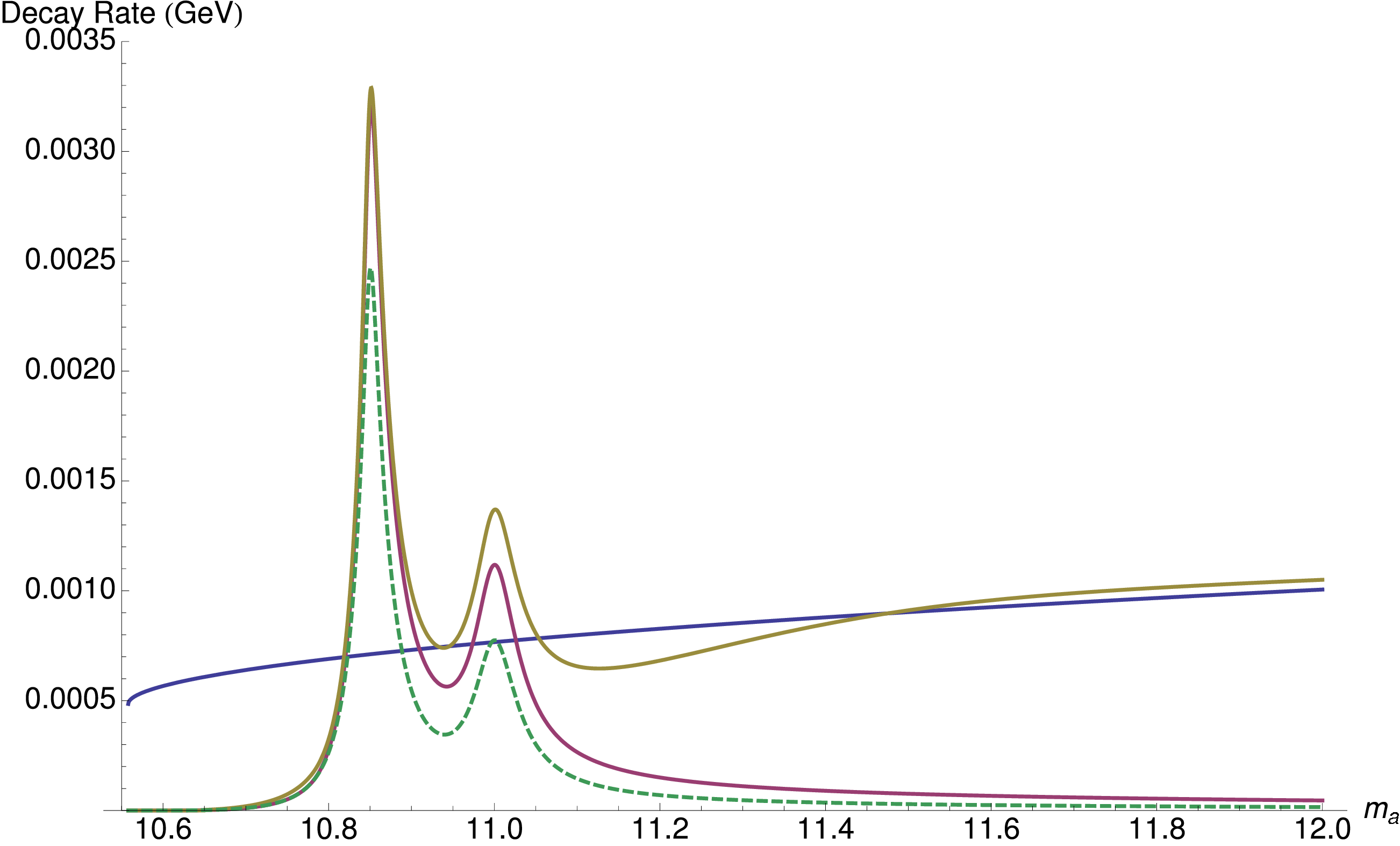} 
\caption{Decay rates in GeV of the $a$ to $b \bar b$ inclusive.  (Blue): Partonic QCD computation assuming $m_b = m_B = 5.279$ GeV with NLO correction given in 
Ref.~\cite{Drees:1989du}.  (Red): Decay rate of $a$ to $b$-flavored mesons from its mixing with $\eta_b$ states.  (Yellow):  Interpolation given by 
Eq.~\ref{eq:interp} to cover region near $B$ meson threshold.  (Green): Mixing calculation without $B_s$ decays, which violate the assumption of uniform coupling, 
$a_{B_s}$ in Eq.~\ref{eq:bmodel} at the order of magnitude level.	}
\label{fig:bbDecays}
\end{figure}

To calculate the $a$ decay rate into the states of Eq.~\ref{eq:bmesonrates}, we replace $\eta_b$ with $a$, and the use the effective coupling on the RHS,
\beqa
|a_{B\,a}|^2 &=& |a_{B5}|^2 |c_5|^2 + |a_{B6}|^2 |c_6|^2 \\
|a_{B_s\,a}|^2 &=& |a_{B_s 5}|^2 |c_5|^2 + |a_{B_s 6}|^2 |c_6|^2 \nonumber, 
\label{eq:mixedcoeffs}
\eeqa
where $c_{5,\,6}$ are given Eq.~\ref{eq:superpos}.  We have made an assumption about the relative phase between $a_{B5}$ and $a_{B6}$, but our final answer will have a small dependence on this as typically only one of $c_5$ or $c_6$ is large.
We plot the inclusive decay rate of $a$ to $b \bar b$ in Fig.~\ref{fig:bbDecays}, seeing two pronounced peaks.  We use the interpolation function of \cite{Drees:1989du}
to take us from the decays due to mixing near threshold to the partonic calculation at larger $m_a$:  
\beq
\Gamma(a \rightarrow b \bar b) = \Gamma(a \rightarrow B,\,B_s \; {\rm mesons\; via}\; \eta_b) + \Gamma_{\rm partonic-NLO}(a \rightarrow b \bar b) \left( 1 - \exp \left[ - \left(\frac{x}{a} \right)^b  \right]  \right),
\label{eq:interp}
\eeq
where $x = \sqrt{1 - \frac{(M_B + M_{B^*})^2}{m^2_a}}$, $b =5$ so that the mixing contribution dominates just above threshold, and $a = 0.32$ is a sum-rule inspired normalization factor so that the 
integral of the interpolating function above threshold equals that of NLO partonic calculation.

%%%%%%%%%%%%%%%%%%%%%%%%%%%%%%%%%%%%%%%%%%%%%%%%%%%%%%%%%%%%%%
% References %%
%%%%%%%%%%%%%%%%%%%%%%%%%%%%%%%%%%%%%%%%%%%%%%%%%%%%%%%%%%%%%%

\bibliography{lit}
\bibliographystyle{apsrev}

\end{document}